# Theory of Electro-Ionic Perturbations at Supported Electrocatalyst Nanoparticles

Yufan Zhang,[1,2] Tobias Binninger,[1] Jun Huang,[1,2] Michael H. Eikerling[1,2]

[1] *Theory and Computation of Energy Materials (IET-3), Institute of Energy Technologies, Forschungszentrum Jülich GmbH, 52425 Jülich, Germany*

[2] *Chair of Theory and Computation of Energy Materials, Faculty of Georesources and Materials Engineering, RWTH Aachen University, 52062 Aachen, Germany*

Nanoscopic heterogeneities in composition and structure are quintessential for the properties of electrocatalyst materials. Here, we present a semiclassical model to study the electrochemical properties of supported electrocatalyst nanoparticles (NP). The model captures the correlated electronic and ionic equilibration across NP, support, and electrolyte. It reveals peculiar trends in surface charging of the supported NP, validated by comparison with first-principles calculations. Support-induced perturbations in electronic and ionic charge densities at the NP's active surface manifest as distinct potentials of zero local electronic and ionic charges that could differ by more than 0.5 V in the studied system.

Electrocatalyst nanoparticles (NPs) anchored on a supporting material are key components in electrochemical devices, such as fuel cells and electrolyzers [1–3]. While the support primarily ensures stable anchoring for NPs, it also impacts their electrocatalytic activity [4–8]. A class of effects, subsumed as metal–support interactions (MSI), have garnered wide attention. However, the physical phenomena underlying these effects remained elusive [9–15].

First-principles calculations based on Kohn-Sham density functional theory (DFT) have elucidated the electron redistribution at metal–support interfaces in vacuum [16–22], corroborated by photoelectron spectroscopy and electron holography [23,24]. This electron redistribution regulates electrochemical properties of surface sites at the particle, and tunes adsorption strengths of reaction intermediates [25–29]. Immersed in an electrolyte solution, physiochemical properties of NP and support co-determine characteristics of the electrochemical double layer (EDL) on the heterogeneous surface. Overlapping EDLs of support and NPs modulate the local reaction environment [30] and, thereby, tune the electrocatalytic activity, as revealed by experiments and rationalized by continuum modelling [31,32].

While previous studies focused on either electronic or ionic MSI effects, it was emphasized that both are coupled through the equilibration across the NP, support, and electrolyte [9]. However, there is a lack of self-consistent treatment for the electronic–ionic coupling at nanoparticle-based electrodes. EDL models based on modified Poisson-Boltzmann (MPB) theory neglect electronic equilibration inside the solid phase [33]. First-principles calculations of electrochemical interfaces face challenges not only in constant-potential simulations [34–38] but also in their considerable computational cost, rendering them unsuitable for systems containing thousands of atoms, such as electrolyte-surrounded supported nanosized particles. Consequently, previous studies of electronic MSI effects were restricted to the contact interface between a sub-nanometer cluster and its support in vacuum [22,39]. It remains unclear in what way electrons redistribute at the external surface of NPs in contact with electrolyte. As a result, there is a notable gap in understanding joint effects of electron redistribution and EDL overlap on local electrochemical properties.

The recently developed semi-classical density-potential functional theory (DPFT) represents a physically consistent approach to treat electronic and ionic degrees of freedom [40–43]. By handling metal electrons through orbital-free DFT [44–46], this approach ensures high efficiency while capturing electronic equilibration and electron spillover phenomena. Furthermore, it enables constant-potential simulations by fixing the electrochemical potential of electrons, $\tilde{\mu}_\text{e}$. Meanwhile, solvents and ions are treated using statistical field theory [47–49].

Application of DPFT to planar silver electrodes has demonstrated its suitability for modelling electrochemical interfaces [43]. Here, we adapt the methodology to study a model system of silver (Ag) NPs supported on a polycrystalline (pc) gold (Au) surface. Metal atomic-core charges are uniformly distributed according to the jellium model. Model parameters for the orbital-free DFT description, metal-solvent interactions, and electrolyte species are calibrated step-by-step, using experimental data for the work function, $\Phi$ [50–54], the potential of zero (free) charge, PZC [55–57], and the differential capacitance, $C_\text{d}$ [55,56] of Ag and Au electrodes, respectively. The orbital-free DFT parameters for Ag are calibrated to the $\Phi$ of the commonly exposed Ag(111) facet ($\Phi = 4.60$ eV). The calculated $\Phi$ for the Ag NP was 4.23 eV, lowered by 0.37 eV due to NP size and surface curvature effects, in good agreement with a 0.33 eV decrease obtained by atomistic DFT simulations (*cf. Supplementary Materials* [58]). In comparison, the Au support has a work function of 5.20 eV, approximately 1 eV larger than that of the Ag NP. Details on calibration procedure, model equations and boundary conditions are provided in *Supplemental Materials*.

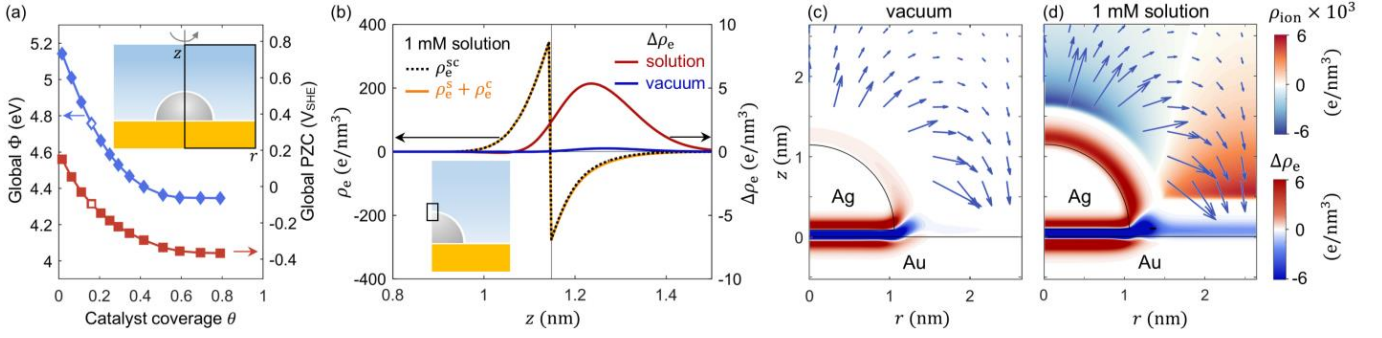

FIG. 1. (a) Global $\Phi$ of Au-supported Ag NP in vacuum and global PZC in 1 mM KClO$_4$ solution as a function of $\theta$. Inset shows the simulation cell. The open symbols represent $\theta = 16\%$, which is further studied in (b-d). (b) $\rho_e$ and $\Delta\rho_e$ profiles along a line through the top surface of the NP ($r = 0$, $0.8$ nm $< z < 1.5$ nm). The NP jellium edge at $z \approx 1.15$ nm is displayed by a vertical line. The orange solid and black dotted curves represent $\rho_e$ before and after contact, respectively, in solution. The red and blue curves denote the $\Delta\rho_e$ in solution and in vacuum, respectively. (c,d) Visualization of $\rho_{ion}$ and $\Delta\rho_e$ for the overall charge-neutral supported NP in vacuum (c) and in solution (d). No ions approach closer than 0.5 nm to the solid surface, leaving a white region therein. The arrows represent electric field.

Specific interactions between an uncharged metal surface and the first layer of solvent molecules determine the interfacial potential distribution and thereby contribute to the shift from $\Phi$ to PZC, as established by experiments [75–77] and first-principles calculations [54,78–83]. Our model captures electrostatic interactions between the spillover electrons and the first-layer water molecules, yielding (i) reduction of interfacial potential drop due to increased dielectric permittivity, and (ii) water dipoles reorientation [61,63–65]. Due to the orbital-free feature of DPFT, electron exchange caused by hybridization of solvents and metal surface is not explicitly modelled [54,79,81]. To address this issue, we apply a heuristic approach by introducing an auxiliary field, $\vec{A}$, which induces a net polarization in the interfacial region. Perpendicular to the surface, $\vec{A}$ is set to decay to zero within a few Angstroms from the surface. Coupled with the solvent dipole moment $\vec{p}$, $\vec{A}$ induces an additional reorientation of solvent dipoles in the first layer, thereby creating an interfacial polarization and yielding the experimental PZC values through tuning the magnitude of $\vec{A}$.

The calibrated model is applied to a hemispherical Ag NP with radius $R_{cat} = 1$ nm, attached to an Au support. The system's axial symmetry reduces the dimensionality to 2D, with radial coordinate $r$ and surface-normal coordinate $z$ (see inset in FIG. 1(a)). The radial size of the simulation cell, $R_{cell}$, determines the fraction of the support surface covered by the NP, *viz.*, $\theta = R_{cat}^2/R_{cell}^2$. The governing equations are solved using COMSOL. The equilibrium distance between the NP basal plane and the support surface has been determined by minimizing the free energy of the total system as a function of the NP–support distance. The minimum occurs at a distance of 0.1 nm (see FIG. S2). As a key property of this study, the local net charge density of the electrode is calculated as $\rho_e = e_0(n_{cc} - n_e)$, where $n_{cc}$ and $n_e$ represent the number densities of metal core charges and electronic charges, respectively. The local net charge density of the electrolyte is determined as $\rho_{ion} = e_0(n_c - n_a)$, with $n_c$ and $n_a$ denoting the number densities of cations and anions, respectively.

Before contact, the Fermi levels of a charge-neutral Ag NP and a charge-neutral Au support in vacuum differ by approximately 1 eV, obtained from the work function difference. Upon contact, while the supported NP system remains overall charge-neutral, electrons redistribute to equilibrate the Fermi levels [16–18]. The energy difference between the equilibrated Fermi level and the vacuum level is the global $\Phi$ of the combined system, which is affected by the $\Phi$ of individual components as well $\theta$. As shown in FIG. 1(a), in the low-coverage limit, the global $\Phi$ approaches 5.20 eV, corresponding to the bare Au support. As $\theta$ increases, a greater portion of the electrode surface becomes occupied by Ag NPs, which have a lower $\Phi$, leading to a decrease in the global $\Phi$.

In 1 mM solution, the global PZC corresponds to the electrode potential, at which the overall $\rho_{ion}$ (or $\rho_e$) in the solution (or electrode) equals zero. FIG. 1(a) demonstrates that both the global PZC and global $\Phi$ exhibit a nonlinear relationship with $\theta$. Notably, the nonlinear shape of these curves is a signature of electron redistribution on the *external surfaces* of the NP and support, a finding first proposed in Ref. [22] and further discussed in *Supplementary Materials*. Herein, we directly test this argument by comparing $\rho_e$ before and after contact in FIG. 1(b) for $\theta = 16\%$ and solution environment. The intrinsic charge separation across the jellium edge of NP at $z \approx 1.15$ nm due to electron spillover is clearly seen, both for the isolated and the supported NP. However, the additional electronic charge induced by Fermi-level equilibration can hardly be discerned on the scale of the intrinsic $\rho_e$, as shown by the overlapping curves of $\rho_e^{sc}$ for the combined supported catalyst (sc) system and $\rho_e^s + \rho_e^c$ for the superposition of support (s) and catalyst (c) systems calculated separately. The electronic perturbation established upon contact can be better represented by the electronic charge density difference between the combined and separate systems,

$$\Delta \rho_e = \rho_e^{sc} - \rho_e^s - \rho_e^c. \tag{1}$$

Positive $\Delta \rho_e$ at the NP surface corresponds to electron depletion, consistent with the direction of electron transfer from Ag NP to Au support. In solution environment, given an atom density of 100 atoms/nm³, the $\Delta \rho_e$ peak value of 5 e/nm³ translates to an increase in atom valency by 0.05 e/atom. Integration of $\Delta \rho_e$ along the $z$-axis yields a considerable net surface charge density of approximately 20 $\mu C/cm^2$. Given a typical double-layer capacitance of 20 $\mu F/cm^2$, the support-induced surface charge perturbation is comparable to the charging caused by an electrode potential shift of about 1 V. This magnitude is sufficient to exert notable catalytic effects such as the modification of activation barrier heights through electric interactions [29] and the shift in adsorbate binding energies by changing the occupancy of adsorbate-related (anti)bonding orbitals [22].

Compared to the solution environment, $\Delta \rho_e$ in vacuum is 20 times smaller. This is caused by the much lower dielectric constant of the medium surrounding the supported NP system. As illustrated in FIG. 1(c,d), a surplus of positive charges (red shaded) locates at the external surface of NP, compensated by an excess of negative charges (blue shaded) at the surrounding support surface. The "built-in" potential difference caused by the Fermi-level alignment generates an electric field surrounding the combined system, pointing from the NP to the support, depicted by arrows. The external surfaces of the NP and the support can be conceptualized as two opposing plates of a capacitor, separated by a dielectric medium (either vacuum or solvent).

$\Delta \rho_e$ at the NP-support contact region ($z \approx 0$, $0\,\text{nm} < r < 1\,\text{nm}$) exhibits a triple-layer structure, in contrast to the conventional double-layer depiction associated with contact charge transfer [17]. The triple-layer structure arises from two superimposed effects: (1) net electron transfer from NP to support and (2) pushback of electron tails at both metal surfaces. Unlike the pure electrostatic models popular in the literature, our model explicitly considers metal electrons and thereby captures electron spillover. For an individual NP or support, the spillover electrons spread to as far as 0.4 nm away from the surface, as shown in FIG. S3. Upon "contact" at a distance of 0.1 nm, these electrons are slightly repelled, moving closer to the metal surfaces, resulting in a net electron excess in the gap region [84].

Although the solution phase maintains globally charge-neutral, variations in the electric potential in solution results in lateral separation of ions, as shown in FIG. 1(d). The blue shades above the NP indicate anion accumulation, while the red shades above the support imply cation excess. In these regions, local ion concentrations (~10 mM) surpass the bulk concentration (1 mM) by an order of magnitude. The gradient from blue through white to red reflects the transition from anion to cation accumulation as one moves in the solution from the NP to the support. This transition results from ionic MSI. Along the heterogeneous surface, EDLs of NP and support overlap, co-determining the electric potential distribution in solution, explaining why the maximal local $\rho_{ion}$ is three orders of magnitude lower than the maximal local $\Delta \rho_e$.

DFT calculations have been performed on a truncated octahedron Ag NP with a height of 1 nm on an Au(111) slab to validate DPFT results. Detailed explanations, along with FIG. S7-S11 and Table S6, are provided in *Supplementary Materials*. The coverage-dependent global $\Phi$ calculated by DFT agrees with DPFT predictions, with deviations of only 0.1 eV. In vacuum, the additional electronic charge densities at the NP's top surface calculated by DPFT and DFT show quantitative agreement. The solvent-induced enhancement of external-surface charging is further supported by DFT calculations with an implicit solvation model. The overall net dipole moment in the simulation cell, resulting from electron redistribution, shows a relative error of 1.8% in vacuum and 5.2% in solution environment when comparing DPFT and DFT calculations. Additionally, a similar triple-layer structure at the contact interface is observed in the DFT calculations.

So far, we have compared the supported catalyst with separate catalyst and support, all in their uncharged states. The constant-charge condition holds relevance in vacuum, gas-phase, or colloidal suspensions in liquid. In the realm of electrocatalysis, however, experiments are performed under controlled electrode potential, $E$. Hence, it is relevant to investigate how the presence of the support perturbs the electro-ionic properties of the NP under controlled $E$. We choose the reference to be an unsupported Ag NP at its own PZC ($E = \text{PZC}_c = -0.52$ V$_{\text{SHE}}$), and focus on the region near the NP top surface ($r = 0$, $0.9\,\text{nm} < z < 2.5\,\text{nm}$), indicated by the black box in the insets of FIG. 2. Given charge-neutrality of the unsupported catalyst NP and thus $\rho_{ion} = \rho_{ion}^c(\text{PZC}_c) = 0$, this choice of reference offers the convenience to represent the support-induced $\rho_{ion}$ perturbations by the actual $\rho_{ion}$ for the supported NP electrode, *viz.*, $\Delta \rho_{ion} = \rho_{ion}^{sc}(\text{PZC}_c) - \rho_{ion}^c(\text{PZC}_c) = \rho_{ion}^{sc}(\text{PZC}_c)$.

Meanwhile, in reference to the $\rho_e$ of the unsupported NP, the support-induced $\rho_e$ perturbation is calculated by

$$\Delta\rho_e = \rho_e^{sc}(\text{PZC}_c) - \rho_e^c(\text{PZC}_c). \tag{2}$$

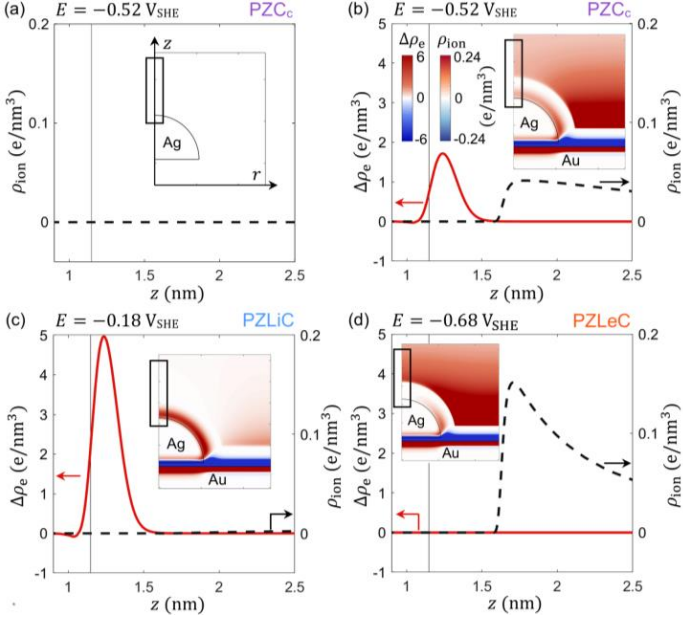

FIG. 2. Impact of Au support on local electronic and ionic charge densities of supported Ag NP under potential control, in 1 mM KClO$_4$ solution and $\theta = 16\%$. The NP jellium edge at $z \approx 1.15$ nm is indicated by a vertical line. (a) Unsupported NP at $E=\text{PZC}_c = -0.52$ V$_{\text{SHE}}$. (b) Supported NP at $E = PZC_c$. (c) Supported NP at $E=\text{PZLiC}= -0.18$ V$_{\text{SHE}}$. (d) Supported NP at $E=\text{PZLeC}= -0.68$ V$_{\text{SHE}}$.

FIG. 2(b) reveals stark support-induced perturbations in the electro-ionic conditions of the NP. The red curve indicates a surplus of positive electronic charges at the top surface, caused by the electronic MSI, similar to the situation in FIG. 1(b). While one might intuitively expect that these excess positive electronic charges would cause an accumulation of anions, we instead observe a cation excess near the NP, denoted by the black dashed curve. How can the positively charged surface attract cations? This seeming contradiction is rationalized by the ionic MSI. The EDLs of the NP and the support overlap at the heterogeneous surface and thereby co-determining the distribution of electric potential. As the applied electrode potential ($-0.52$ V$_{\text{SHE}}$) is more negative than the PZC of the Au support (0.19 V$_{\text{SHE}}$), the strongly negatively charged support creates a negative potential region that extends above the support and the NP. This is confirmed by the $\phi$ profile along a path located 0.5 nm above the electrode, illustrated by the purple curve in FIG. 3(a). Consequently, cations accumulate around the NP, despite its net positive electronic charges. To summarize, although global electroneutrality in the simulation cell always holds, locally at the NP top, electronic and ionic excess charges *of the same polarity* are generated by the coupled electro-ionic MSI. This surprising finding constitutes a major result of the study.

For a homogeneous electrode, the PZC is the characteristic electrode potential corresponding to the charging state of zero electronic and ionic excess charges. Because of the support-induced electro-ionic perturbations, the PZC$_c$ of the unsupported NP ($-0.52$ V$_{\text{SHE}}$) no longer guarantees zero excess charges of the supported NP and thereby fails to indicate its charging state. It is natural to ask whether a new characteristic electrode potential for the supported NP, akin to the PZC$_c$ for unsupported NP, can be identified. To restore zero $\rho_{\text{ion}}$ above the NP surface, corresponding to $\phi = 0$ in that region (see the blue curve in FIG. 3(a)), $E$ must be shifted positively by 0.34 V relative to the PZC$_c$, as demonstrated in FIG. 2(c). The resulting $E$ is defined as the "potential of zero *local ionic* charge" (PZLiC). However, despite the restoration of the local $\rho_{\text{ion}}$, the local $\rho_e$ deviates further from the reference value of the unsupported charge-neutral NP, as revealed by plotting $\Delta\rho_e = \rho_e^{sc}(E) - \rho_e^c(\text{PZC}_c)$ in FIG. 2(c). In contrast, to restore the local $\rho_e$ at the NP top surface, $E$ must be negatively shifted in opposite direction by $-0.16$ V relative to the PZC$_c$, as shown in FIG. 2(d). Analogous to the PZLiC, we term this $E$ as the "potential of zero *local electronic* charge" (PZLeC).

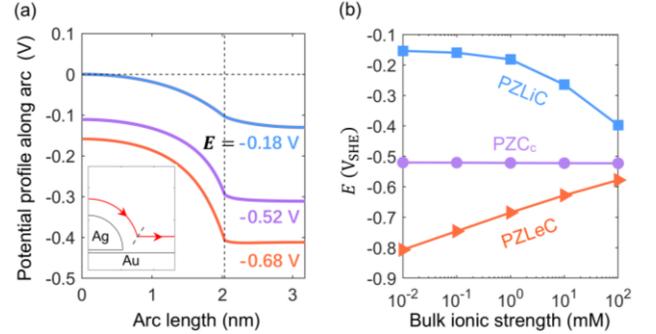

FIG. 3. (a) Potential profiles along a path located 0.5 nm from the electrode surface (see the red curve in the inset). Each profile is calculated at the electrode potential indicated nearby. (b) Variation of the PZLiC and PZLeC of supported NP as well as the PZC$_c$ of unsupported NP with bulk ionic strength.

While at PZC$_c$ ($-0.52$ V$_{\text{SHE}}$), the unsupported catalyst NP exhibits both zero electronic and ionic excess charges, we have demonstrated that no single $E$ exists for the supported NP that ensures the absence of both electronic and ionic excess charges. Moreover, the PZLiC and PZLeC deviate from the PZC$_c$ in opposite directions, with a difference between them reaching up to 0.5 V. Consequently, the support-induced electro-ionic perturbations yield *qualitatively* different electrochemical conditions for catalyst NPs, that are expected to strongly influence their electrocatalytic properties.

The electro-ionic perturbations around the NP are most pronounced in dilute electrolyte. In the case of 1 mM concentration (FIG. 2(b)), $\rho_e$ at the NP top alters by approximately 2 e/nm$^3$ and the cation concentration nearby increases by a factor of 70 compared to the bulk solution. At higher ionic strength (100 mM, FIG. S4.), these values decrease to around 0.5 e/nm³ and a factor of 2. FIG. 3(b) shows the variation of the three characteristic potentials with electrolyte concentration. The PZLiC and PZLeC converge to the $PZC_c$ at high concentration, a result of the electro-ionic decoupling between the NP and support at increased ionic screening. Additionally, the size dependency is shown in FIG. S5, where the deviation of the PZLeC from the $PZC_c$ decreases from 0.29 V for a 1 nm radius NP to 0.04 V for a 4 nm radius NP.

The present modelling framework offers insights into local electro-ionic conditions that are challenging to access experimentally. It is, of course, essential to provide experimentally attainable predictions. We therefore calculated $C_d$ of the supported NP electrode. While capacitance features of homogeneous planar electrodes have been extensively studied, much less is known about those of composite electrodes. The prevailing view on the $C_d$ of heterogeneous surfaces suggests a linear superposition of $C_d$ of individual constituents. For example, the $C_d$ of a polycrystalline electrode was theoretically calculated as a surface-area-weighted average of constituting facets [85–88]. Similarly, the capacitive response of a composite supported NP electrode is considered to be a blend of the responses from the individual NP and support, manifesting as an averaging effect—the simplest form being a linear superposition weighted by their respective exposed surface areas.

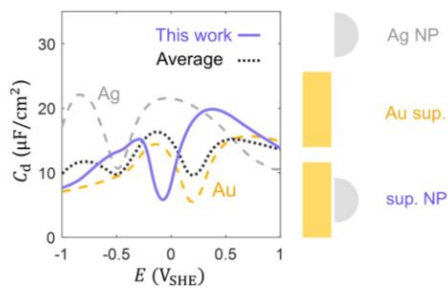

FIG. 4. $C_d$ of Au-supported Ag NP electrode at $\theta = 16\%$ in 1 mM KClO₄ solution (solid line). The NP has a radius of 1 nm. The dashed lines depict the $C_d$ of Ag NP and planar pc Au electrodes, while the dotted line illustrates a surface-area-weighted average of the two reference curves.

FIG. 4 shows the calculated $C_d$ of the supported NP system in a 1 mM solution (purple line). Reference curves for the Ag NP and Au support are plotted as silver and gold dashed lines, respectively, each displaying a minimum at its PZC. The linear superposition of these two lines, depicted by the black dotted line, exhibits two minima inherited from the two constituents. In contrast, the $C_d$ curve for the supported NP from our model reveals only a single minimum located at the *global* PZC, as capacitance measurements capture only the total charge flowing through the external circuit, without distinguishing the specific locations where the charges are localized. This single-minimum feature sharply contrasts with the linear superposition assumption, implying the electro-ionic coupling that homogenizes the two entities' capacitive response.

For larger NPs, however, the electro-ionic coupling weakens. As shown in FIG. S6, the $C_d$ curve of supported NP electrodes gradually develops three peaks and two local minima with increasing NP size, resembling the superposition of the capacitance contributions from individual components.

In conclusion, we have theoretically investigated the local electro-ionic conditions of supported NP electrodes. The exemplary case of a silver NP attached to a gold support reveals strong support-induced perturbations of electron and ion densities at the electrocatalyst surface. These findings demonstrate that electronic and ionic effects are strongly coupled in supported electrocatalyst systems. They are best described as joint *electro–ionic* metal–support interactions (EIMSI).

The authors acknowledge funding from the Helmholtz Young Investigators Group program and the Helmholtz-Gemeinschaft Deutscher Forschungszentren e.V. (HGF), Program-oriented Funding (PoF IV), under the Research Program: Materials and Technologies for the Energy Transition (MTET). The authors also acknowledge the financial support from the Horizon Europe project DECODE founded by the European Union under the Grant Agreement number 101084131. Y.Z. is grateful to Xinwei Zhu and Dr. Yuankai Yang from Forschungszentrum Jülich, Jinwen Liu from Leiden University, Dr. Qing Wang from Epma and Dr. Mengzi Huang from ETH Zürich for helpful discussions. Y.Z. also thanks Dr. Weiqiang Tang and Dr. Piotr Kowalski from Forschungszentrum Jülich, Dr. Leyu Liu and Prof. Hai Xiao from Tsinghua University for their guidance on DFT calculations. Special thanks are extended to Xiaoyue Wang from WTW for her valuable suggestions on the color scheme used in the figures. Finally, the authors would like to acknowledge anonymous reviewers for valuable insights that improve the quality of the paper.